\documentclass[submission,copyright,creativecommons,sharealike,noncommercial]{eptcs}
\providecommand{\event}{QPL 2019}
\pdfoutput=1
\usepackage{mathtools} 
\usepackage{amstext} 
\usepackage{amssymb} 
\usepackage{bbold} 
\usepackage{amsthm} 
\usepackage{stmaryrd} 

\usepackage[utf8]{inputenc} 

\usepackage{relsize} 
\usepackage{microtype} 
\usepackage{csquotes} 
\usepackage{ragged2e} 

\usepackage{hyperref} 
\usepackage[compress]{cite} 

\usepackage{graphicx} 
\usepackage[usenames,dvipsnames]{xcolor} 
\usepackage{stackengine}
\usepackage{tikz} 
\usepackage{circuitikz} 
\usetikzlibrary{
	arrows,
	shapes,
	decorations,
	intersections,
	backgrounds,
	positioning,
	circuits.ee.IEC
	}

		\newcounter{theorem_c} 
		\numberwithin{theorem_c}{section} 
		\numberwithin{equation}{section} 

		\theoremstyle{plain}
		\newtheorem{theorem}[theorem_c]{Theorem}
  \newtheorem{proposition}[theorem_c]{Proposition}

		\newtheorem{corollary}[theorem_c]{Corollary}

		\newtheoremstyle{exampstyle}
		  {2mm} 
		  {2mm} 
		  {\itshape} 
		  {} 
		  {\bfseries} 
		  {.} 
		  {.5em} 
		  {} 

		\theoremstyle{exampstyle}
		\newtheorem{definition}[theorem_c]{Definition}
		
		\newtheorem{remark}[theorem_c]{Remark}

\usepackage{tikzit}

\tikzstyle{green dot}=[fill=green, draw=black, shape=circle]
\tikzstyle{red dot}=[fill=red, draw=black, shape=circle]
\tikzstyle{smallbox}=[fill=white, draw=black, shape=rectangle, minimum width=0.2cm, minimum height=0.2cm]
\tikzstyle{box}=[fill=white, draw=black, shape=rectangle, minimum width=1cm, minimum height=1cm]
\tikzstyle{widebox}=[fill=white, draw=black, shape=rectangle, minimum width=2cm, minimum height=1cm]
\tikzstyle{tallbox}=[fill=white, draw=black, shape=rectangle, minimum width=1cm, minimum height=2cm]

\newcommand{\suchthat}[2]{\left\{\, #1 \,\middle\vert\, #2 \,\right\}}
\newcommand{\Forall}[2]{\forall\,#1.\;#2}
\newcommand{\Exists}[2]{\exists\,#1.\;#2}

\newcommand{\orthosubgroups}[2]{#1 \bot #2}
\newcommand{\Stab}[3]{\operatorname{Stab}^{#1}_{#2}\left(#3\right)}

\newcommand{\States}[1]{\operatorname{St}\left(#1\right)}
\newcommand{\StatesPure}[1]{\operatorname{St}\left(#1\right)_{\text{pure}}}
\newcommand{\Transf}[1]{\operatorname{Transf}\left(#1\right)}
\newcommand{\Restr}[2]{\left[#2\right]_{#1}}
\newcommand{\ProcessTheory}[2]{\operatorname{PT}\left(#1,#2\right)}
\newcommand{\ProcessTheoryPure}[2]{\operatorname{PT}\left(#1,#2\right)_{\text{pure}}}

\newcommand{\picturescaling}{0.8}

\title{A Process-Theoretic Church of the Larger Hilbert Space}

\author{
	Stefano Gogioso\\
	University of Oxford \\
	\texttt{stefano.gogioso@cs.ox.ac.uk}
}
\date{}

\def\titlerunning{A Process-Theoretic Church of the Larger Hilbert Space}
\def\authorrunning{Stefano Gogioso}

\begin{document}
\maketitle

\begin{abstract}
	We show how to reconstruct a process theory of local systems starting from a global theory of reversible processes on a single global system, by using the purification principle. In such a process theory, local systems are not given, but rather ``emerge'' as the global system is decomposed into subsystems. Local systems thus have specific identities and their composition is naturally limited by structural constraints, a behaviour which we formalise by defining symmetric partially-monoidal categories. We reconstruct quantum theory from the global theories of unitary groups acting on projective Hilbert spaces.
\end{abstract}

\section{Introduction}
\label{section:introduction}

The ``Church of the Larger Hilbert Space'' is a an expression---originally attributed to John Smolin---used to denote the interpretation of quantum states and channels as arising from global pure states and unitary operations by the discarding of environments. From a purely mathematical perspective, the ``Church of the Larger Hilbert space'' is nothing more than an incarnation of Stinespring's Dilation. From an operational and process-theoretic perspective \cite{abramsky2004categorical,abramsky2009,selinger2011,coecke2013causal,PhysRevA.84.012311,PhysRevA.81.062348,chiribella2014dilation}, however, it denotes a philosophical stance on the epistemic nature of local states and processes, interpreted to arise from the ontic global states and processes because of the local inaccessibility of certain systems (the \emph{environments}). It can be interpreted to say that states and processes should emerge from a global perspective by decomposition, rather than being defined locally and subsequently composed.

Global approaches to quantum theory have flourished since its early days---dating all the way back to Schr\"{o}dinger's work, through Everett's PhD thesis \cite{everett1956theory} and later applications to general relativity \cite{hartle1983wave} and quantum field theory \cite{haag1964algebraic,halvorson2006algebraic}, to only mention a few---but have only recently been explored from an abstract process-theoretic perspective. In \cite{delrio2017operational}, the authors describe the condition of no-signalling between distant agents in terms of commutants in a generic monoid of global processes, define local states as equivalence classes and study the general implications on secrecy. In \cite{chiribella2018agents}, the author generalises the previous approach, defining both states and transformations for sub-systems. The work is framed within the context of the purification principle and many concrete examples of sub-systems in theories of interest to quantum information are worked out, showing the definition to have wide applicability.

Both previous works, however, share a key deficiency: the fundamental ingredient common to virually all process-theoretic approaches is the \emph{compositional} perspective, and yet the monoidal structure of emergent systems\footnote{A term which we broadly use to mean ``local systems'', ``sub-systems'', ``maximal agents'', etc.} is not recovered. As this work will show, \emph{defining} systems is the easy part of the decompositional story: the hard part, where all the complexities lie, is describing how systems can be made to \emph{compose}. Unlike previous approaches to the decomposition of systems into tensor products \cite{zanardi2004tensor}---which exploited the structure of measurements and observables---the compatibility conditions we derive are purely group-theoretical and apply to arbitrary theories of global reversible processes.

\section{Reversible global process theories}
\label{section:theories}

As our starting point, we consider a generic theory of global reversible processes. In first instance, we only know two things about such a theory: its global states and the group of global transformations $\Xi$ acting on said states. In the spirit of the Erlangen program, we restrict ourselves to theories where global states form a homogeneous space $\Theta$ under the action of global transformations.
\begin{definition}
	A \emph{reversible global process theory} is a homogeneous space $\Theta$ for a centreless group $\Xi$, i.e. a transitive representation\footnote{By a \emph{representation} of a group $\Xi$ we will mean a group homomorphism $\Xi \rightarrow \operatorname{Aut}(\Theta)$ of $\Xi$ into the automorphisms of an object $\Theta$ of some concrete category. We will not require any more structure than a group action on a set $\Theta$ in this work, but additional structure may be imposed by appropriate choice of category, if needed.} of $\Xi$. We refer to $\Xi$ as the \emph{group of global transformations} and to the space $\Theta$ that it acts upon as the \emph{space of global states}. For each \emph{global state} $\psi \in \Theta$, we refer to the stabilizer $\Stab{\Theta}{\Xi}{\psi}$ as the \emph{phase group} for the state.
\end{definition}
\noindent As a reminder, the \emph{centre} $Z(\Xi)$ of a group $\Xi$ is the normal subgroup formed by those elements $\xi \in \Xi$ which commute with all other elements of $\Xi$, i.e. $Z(\Xi) := \suchthat{\xi \in \Xi}{\Forall{\chi \in \Xi}{\xi \chi = \chi \xi}}$. A group $\Xi$ is said to be \emph{centreless} if its centre is trivial, i.e. if $Z(\Xi) = \{1\}$.

In a process theory, reversible transformations $h: H \rightarrow H$ and $k: K \rightarrow K$ on two systems $H$ and $K$ can be canonically interpreted as reversible transformations $h \otimes 1_K$ and $1_H \otimes K$ on the same joint system $H \otimes K$, and as such they always necessarily commute.
Starting from global transformations, without any prior knowledge of systems, the compositional picture is not immediately available and one has to instead resort to what we shall refer to as a \emph{decompositional} approach. If one is given subgroups $H,K \leq \Xi$ of transformations acting locally on some systems $\mathcal{H}$ and $\mathcal{K}$, then under the identification above it is possible to consider the tensor product $\mathcal{H} \otimes \mathcal{K}$ only if all transformations in $H$ always commute with all transformations in $K$. In this work, we will not interpret local systems as arbitrary ``actors'', but instead we will want the transformations on a system to reflect the maximal set which is physically compatible with any given environment: this means that the transformations $H$ on a system $\mathcal{H}$ as above should contain \emph{all} those transformations which are compatible with \emph{all} possible choices of $K$ and $\mathcal{K}$ for which a tensor product $\mathcal{H} \otimes \mathcal{K}$ can be defined. In group-theoretic terms, this means that $H$ should be a \emph{self-bicommutant} subgroup of $\Xi$, as defined and investigated in the remainder of this Section.

\begin{remark}
	In this work, the word \emph{transformation} is used to denote the reversible transformations only, arising as subgroups of the global transformations. From a process-theoretic perspective, transformations will be special instances of \emph{processes}, but not all processes will be transformations. This marks a difference with previous literature, where the word transformation is used to denote all processes.
\end{remark}

Given a subgroup $H \leq \Xi$, the \emph{commutant} $H'$ is defined to be the subgroup $H' \leq \Xi$ formed by those elements which commute with all elements of $H$:
\[
	H' = \suchthat{\xi \in \Xi}{\Forall{h \in H}{\xi h = h \xi}}
\]
A subgroup $H \leq \Xi$ is called \emph{self-bicommutant} if it satisfies $H = H''$. The \emph{centre} $Z(H)$ of a subgroup $H \leq \Xi$ is defined as the group of all $h \in H$ which commute with all elements of $H$, a definition which can be written succinctly as $Z(H) := H \cap H'$. In particular, a subgroup $H$ is centreless if and only if $H \cap H' = \{1\}$. Given two self-bicommutant subgroups $H$ and $K$, we can define the \emph{join} $H \vee K$ as follows:
\[
	H \vee K := (H \cup K)'' = (H' \cap K')'
\]
where we have used the identity $(H \cup K)' = (H' \cap K')$. The \emph{meet} $H \wedge K$ is simply defined as the set-theoretic intersection:
\[
	H \wedge K :=  H \cap K
\]
Note that the join is explicitly defined to be self-bicommutant, while the meet can easily be verified to be self-bicommutant using the identity above.
Join and meet of self-bicommutant subgroups are commutative and associative.
In particular, the entire group $\Xi$ and its centre provide units to the join and meet respectively, so that self-bicommutant subgroups form a bounded lattice with bottom element $Z(\Xi)=\{1\}$ and top element $\Xi$. We can also easily check that:
\[
	H \vee K = (H'' \vee K'')'' = (H' \wedge K')'
	\hspace{2cm}
	H \wedge K = H \cap K =(H' \cup K')' = (H' \cup K')''' = (H' \vee K')'
\]
so that \emph{De Morgan laws} are satisfied by the lattice (even though the lattice is not necessarily distributive).

\begin{definition}
	Two self-bicommutant subgroups $H,K$ of $\Xi$ are \emph{orthogonal}, written $\orthosubgroups{H}{K}$, if we have $K \leq H'$ (or equivalently $H \leq K'$).
\end{definition}
Every self-bicommutant subgroup is always orthogonal to its commutant, i.e. we always have $\orthosubgroups{H}{H'}$. We will refer to the self-bicommutant subgroups $H$ further satisfying $H \wedge H' = Z(\Xi) = \{1\}$  as \emph{orthocomplemented}, because they are exactly the orthocomplemented elements of the lattice (with respect to the operation of taking the commutant):
\[
	H \wedge H' = Z(H) = Z(\Xi) = \{1\}
	\hspace{17mm}
	H \vee H' = (H \cup H')'' = (H' \cap H'')' = Z(\Xi)' = \Xi
\]
In order to establish a logical parallel with quantum theory, one might also be interested to know for which $H \leq K$ the \emph{orthomodular law} is satisfied:
\[
	H \vee (H' \wedge K) = K
\]
The orthomodular law turns out to be equivalent to the condition that $K$ contains ``enough'' transformations which are a product over $H$, in a sense that will be made clear shortly.

\subsection{Tensor product of transformations}
\label{section:theories/tensor-product-transformations}

Firstly, note that for any two self-bicommutant subgroups $H$ and $K$ we can define the following set:
\[
	H \cdot K = \suchthat{hk}{h \in H, k \in K} \subseteq \Xi
\]
and it is an easy check that $H \vee K = (H \cdot K)''$ as sets. If $H$ and $K$ are orthogonal, then $H \cdot K$ is in fact a subgroup of $\Xi$, a fact which we can use to show that the join between two orthogonal self-bicommutant subgroups behaves as one would expect from a tensor product.

\begin{proposition}
	\label{proposition:tensor-product-transformations}
	Let $H,K \leq \Xi$ be two orthogonal self-bicommutant subgroups and write $H \otimes K := H \vee K$ for their join. Given two transformations $h \in H$ and $k \in K$, define $h \otimes k := hk \in H \cdot K \leq H \otimes K$. The operation $\otimes$ is associative and commutative on both subgroups and transformations, with $Z(\Xi) = \{1\}$ as a bilateral unit on subgroups. Given transformations $a,b \in H$ and $c,d \in K$, the operation $\otimes$ satisfies the following \emph{exchange law}:
	\[
		(ab) \otimes (cd) = (a \otimes c)(b \otimes d)
	\]
	Furthermore, the identity map on $H \otimes K$ is an isomorphism of groups $\sigma_{H,K}: H \otimes K \rightarrow K \otimes H$ which restricts to an isomorphism $\sigma_{H,K}\vert_{H \cdot K}: H \cdot K \rightarrow K \cdot H$ such that $\sigma_{H,K}(h \otimes k) = k \otimes h$ for all $h \otimes k \in H\cdot K$.
\end{proposition}

\noindent The result above states that, at least morally, the operation $\otimes$ is a symmetric monoidal product, so that we are allowed to interpret the transformations $hk \in H \cdot K$ as the transformations on $H \otimes K$ in product form.
\vspace{1mm}
If $H$ is an orthocomplemented system then we are also morally justified to interpret the transformations $h \in H$ as the global transformations $h \otimes 1 \in H \otimes H' = \Xi$ which act trivially ``outside of'' $H$ (i.e. on its orthocomplement $H'$).
In fact, this is the exactly the same as saying that the ``local'' transformations in $H$ should be the equivalence classes of global transformations in $H \cdot H'$ under the action of transformations in $H'$, since $h \otimes 1$ can always be chosen as a canonical representative for each class.

If $H$ and $K$ are orthogonal, then we always have that $Z(H \cdot K) = Z(H) \wedge Z(K) = H \wedge K$, so that the following is an isomorphism of groups:
\[
	(H \cdot K)/Z(H \cdot K) \cong H/Z(H \cdot K) \times K/Z(H \cdot K)
\]
This means that the representation of a transformation in $H \cdot K$ as $hk$ for $h \in H$ and $k \in K$ is unique up to an element of the centre $Z(H \cdot K)$, which we will later show to act trivially on states. If $H$ and $K$ are orthogonal and at least one is orthocomplemented, then $Z(H \cdot K) = Z(H) \wedge Z(K) = \{1\}$ and the representation of a transformation in $H \cdot K$ as $hk$ for $h \in H$ and $k \in K$ is actually unique.

We can now look back to the orthomodular law, which we re-phrase in the following way for all self-bicommutant $H$ and $K$ such that $H \leq K$:
\[
	\big(H \cdot (H'\wedge K)\big)'' = H \vee (H'\wedge K) = K
\]
Given our observations above, we can interpret the subgroup $H \cdot (H'\wedge K) \leq K$ as formed by those transformations in $K$ which factor as a product $k = h \otimes h'$ over the subgroup $H$ and its commutant \emph{within $K$}, the subgroup $H'^{_K} := H'\wedge K \leq K$. The orthomodular law can then be reinterpreted as the statement that there are enough such ``product transformations'' that the only thing in $K$ commuting with all of them is the centre $Z(K)$. In particular, the following instance of the orthomodular law always holds:
\[
	H \vee H'^{_{H \otimes K}} = H \vee (H' \wedge (H \otimes K)) = H \otimes K
\]
It is not true, in general, that $K = H' \wedge (H \otimes K)$ and $H = K' \wedge (H \otimes K)$, though both conditions are themselves instances of the orthomodular law when $\orthosubgroups{H}{K}$ (i.e. when $H \leq K'$ and $K \leq H'$):
\[
	K = H' \wedge (H \otimes K)
	\hspace{2mm} \Leftrightarrow \hspace{2mm}
	K' = H \vee (H' \wedge K')
\]
\[
	H = K' \wedge (H \otimes K)
	\hspace{2mm} \Leftrightarrow \hspace{2mm}
	H' = K \vee (K' \wedge H')
\]

\section{Systems}
\label{section:systems}

In the previous Section, we have identified the subgroups of local transformations and shown that orthogonal subgroups can be composed according to an operation which resembles a tensor product. Local transformations, however, are only half the definition of systems: we still need to define local states and the action of local transformations on them.

In quantum theory, the \emph{purification principle} states that every state $\rho$ for a quantum system $\mathcal{H}$ can be written as the restriction $\rho = \Restr{\mathcal{H}}{\psi}$ of a pure state $\psi$ for a larger quantum system $\mathcal{H} \otimes \mathcal{E}$, and that any two such restrictions $\Restr{\mathcal{H}}{\psi} and \Restr{\mathcal{H}}{\varphi}$ are equal if and only if there is some unitary transformation $v$ on $\mathcal{E}$ such that $\varphi = (1 \otimes v)\psi$. In the more general context of process theories, we could think of the purification principle as saying that every state $\rho$ on a system $\mathcal{H}$ should be obtainable as the restriction $\Restr{\mathcal{H}}{\varphi}$ of some pure state $\varphi$ on a larger system $\mathcal{H} \otimes \mathcal{E}$ and that two such restrictions $\Restr{\mathcal{H}}{\psi}$ and $\Restr{\mathcal{H}}{\varphi}$ should be equal if and only if there is some transformation $v$ on $\mathcal{E}$ such that $\varphi = (1 \otimes v)\psi$. Pure states are then exactly those which purify to product states.

A direct adaptation of the ``compositional'' purification principle above to our decompositional setting might work for orthocomplemented self-bicommutant subgroups $H$: for them we have $H \otimes H' = \Xi$, and the states acted upon by $H$ can arise directly from global states by tracing away the ``maximal environment'' specified by the transformations in $H'$. It does, however, run into a severe issue when the self-bicommutant subgroup $H$ for which we wish to define states is not orthocomplemented: no environment $E$ exists such that $H \otimes E = \Xi$, so that local states acted upon by $H$ cannot be taken to arise from global states by tracing out one subsystem of a tensor product. Contrary to what it may seem, though, this is not a failure of the purification principle itself: it may well be that an environment $E$ could exists such that all states acted upon by $H$ arise by restriction from pure states acted upon by $H \otimes E$. The real issue here is that if we cannot ``reach'' the global system by tensor product then we have not states to start with!

Because of limitations described above, we let go of the ``compositional'' part of the purification principle---the one stating that the purification must be achieved by \emph{tensor product} with an environment---and we define our local states as the equivalence classes obtained by tracing away the commutant $H'$ even in those cases where $H \otimes H' \neq \Xi$. The necessity for such an extension of the purification principle was already discussed in \cite{chiribella2018agents} and our definition of restriction of states---at least when seen as set-theoretic equivalence classes---coincides with the one given therein.

\begin{definition}
	Let $(\Xi,\Theta)$ be a global process theory and $\psi \in \Theta$ be a global state. If $H \leq \Xi$ is a self-bicommutant subgroup, the \emph{restriction} of $\psi$ to $H$ is defined as the following $H'$-orbit in $\Theta$:
	\[
		\Restr{H}{\psi} := H' \psi
	\]
	We write $\States{H} := \Theta/H'$ and refer to its elements as the \emph{local states compatible with $H$}.
\end{definition}
\begin{proposition}
	\label{proposition:systems-action}
	Let $(\Xi,\Theta)$ be a global process theory and $H \leq \Xi$ be a self-bicommutant subgroup. Then $H$ acts on the space $\States{H} := \Theta/H'$ by $h(\Restr{H}{\psi}) := \Restr{H}{h\psi}$. Furthermore, the centre $Z(H)$ always acts trivially, so that the representation $(H,\States{H})$ always descends to a quotient representation $(H/Z(H),\States{H})$.
\end{proposition}

\begin{definition}
	Let $(\Xi,\Theta)$ be a global process theory, $\psi \in \Theta$ be a global state and $H,K \leq \Xi$ be self-bicommutant subgroups such that $K \leq H$. We define the \emph{iterated restriction} of $\rho := \Restr{H}{\psi} \in \States{H}$ to $K$ as follows:
	\[
		\Restr{K}{\rho} := \Restr{K}{\psi} \in \States{K}
	\]
\end{definition}
\begin{proposition}
	\label{proposition:iterated-restriction-well-defined}
	Let $(\Xi,\Theta)$ be a global process theory and $H,K \leq \Xi$ be self-bicommutant subgroups such that $K \leq H$. Write $\rho:= \Restr{H}{\psi}$ for a global state $\psi \in \Theta$. If $ \varphi \in \Theta$ is any global state such that $\Restr{H}{\psi} = \Restr{H}{\varphi}$, then $\Restr{K}{\psi} = \Restr{K}{\varphi}$, so that iterated restriction is well-defined. If $H = \Xi$, then we can canonically identify $\Restr{H}{\psi}$ with $ \psi$ and we have $\Restr{K}{\Restr{H}{\psi}} = \Restr{K}{\psi}$, so that iterated restriction coincides with ordinary restriction when applied to global states.
\end{proposition}

Even though a self-bicommutant subgroup $H$ with its action on $\Theta/H'$ might be taken to constitute a system, we shall not adopt that definition here: the space $\Theta/H'$ of all restricted states is ``too large'', including both pure states---which exist without knowing anything else than a system---and mixed states---which instead require existence and control of environments. The reason why this is relevant marks a major difference in interpretation between the theories of processes obtained through our decompositional approach and process theories in the more standard compositional approaches.

When process theories are identified with (strict) symmetric monoidal categories, the interpretation is that objects in the category define \emph{types} of physical systems, rather than individual physical systems. When one considers the tensor product $\mathbb{C}^2 \otimes \mathbb{C}^2$, one is thinking of the joint system constituted by taking \emph{some system} which has the type of a qubit together with \emph{some other system} which has the type of qubit, without particular concern for the actual identity of the systems themselves. In our decompositional approach, on the other hand, one considers a global theory from which a fixed set of systems will emerge, each one with its unique set of local states and its self-bicommutant subgroup of local transformations. Because each system has its own identity and its own relationship with other systems, the abstract mathematical operation of taking the tensor product must now include an additional requirement for the systems involved to be compatible, in a suitable sense. A similar approach, resulting in a partially defined tensor product, also appears in some of the literature on causality in process theories \cite{coecke2013causal,coecke2016terminality}. In this work, partiality of tensor product will be formalised through the definition of symmetric partially-monoidal categories (closely related to those defined in \cite{coecke2013causal}, see Appendix \ref{appendix:partially-monoidal-categories}).

When it comes to local transformations, previous literature \cite{chiribella2018agents,delrio2017operational} has formalised compatibility of systems as the requirement for the relevant self-bicommutant subgroups to be orthogonal: in this work, we will show that a stronger requirement of ``orthocomplementarity'' is actually needed for compatible systems to compose. Local states further introduce additional requirements: for two mixed states to admit a tensor product, for example, one needs to ask for independence of both the relevant systems and the environments necessary to purify the mixed states. In a partially-monoidal category, definition of a tensor product is a property of systems, not of their respective states and processes: if the tensor product of two systems can be taken, than so can the tensor product of any pair of states or processes. This means that, when we define our process theory, our systems will need to keep track of available environment, precisely because of mixed states. That said, mixed states will arise later on as part of the construction of processes, so we will now focus our attention on pure states only.
\begin{definition}
	Let $(\Xi,\Theta)$ be a global process theory and $H \leq \Xi$ be a self-bicommutant subgroup. We say that a global state $\psi \in \Theta$ is a \emph{product state} over $H$ (eq'tly over $H'$) if the action of $H \cdot H'$ on the orbit $(H \cdot H')\psi$ factors into a product of the action of $H$ on the orbit $H \Restr{H}{\psi}$ and of $H'$ on the orbit $H' \Restr{H'}{\psi}$.
\end{definition}

\noindent In particular, if $\psi$ is a product state over $H$ then we have the following equation:
\[
	\Stab{\Theta}{H\cdot H'}{\psi} = \Stab{\Theta}{H}{\psi} \cdot \Stab{\Theta}{H'}{\psi}
\]
where by $\Stab{\Theta}{G}{\psi}$ we mean the stabiliser of $\psi \in \Theta$ under the action of a subgroup $G \leq \Xi$. Note that the three stabilisers in the equation above would not change if we took them inside the orbit $(H \cdot H')\psi$.

\begin{definition}
	Let $(\Xi,\Theta)$ be a global process theory and $H \leq \Xi$ be a self-bicommutant subgroup. The \emph{pure} local states compatible with $H$ are the restrictions $\Restr{H}{\psi} \in \States{H}$ of the global states $\psi \in \Theta$ which are product over $H$. We write $\StatesPure{H} \subseteq \States{H}$ for the subset of pure local states.
\end{definition}
\begin{proposition}
	\label{proposition:pure-local-states-representation}
	Let $(\Xi,\Theta)$ be a global process theory and $H \leq \Xi$ be a self-bicommutant subgroup. If $\Restr{H}{\psi} \in \States{H}$ is pure then so is $h(\Restr{H}{\psi})$ for all $h \in H$, so that the pure local states compatible with $H$ form a sub-representation $(H,\StatesPure{H})$ of the representation $(H,\States{H})$. Furthermore, the stabiliser of a pure state $\Restr{H}{\psi}$ under the action of $H$ is given by restriction:
	\[
		\Stab{\StatesPure{H}}{H}{\Restr{H}{\psi}} = \Stab{\Theta}{\Xi}{\psi} \cap H
	\]
	In particular, the centre of $H$ acts trivially on pure states.
\end{proposition}

\begin{corollary}
	\label{corollary:states-with-pure-marginals-are-product}
	Let $(\Xi,\Theta)$ be a global process theory and $H \leq \Xi$ be a self-bicommutant subgroup. Consider global states $\psi, \varphi \in \Theta$ and assume that $\psi$ is a product over $H$. If $H'\psi = H' \varphi$, then $\varphi$ is also a product over $H$. Hence whether $\Restr{H}{\psi}$ defines a pure local state compatible with $H$ does not depend on the particular choice of $\psi$.
\end{corollary}

Now that we have defined pure local states and shown that they are well-behaved under the action of self-bicommutant subgroups, we may be tempted to define systems as the representations $(H,\StatesPure{H})$. There are, however, two obstacles to this approach:
\begin{enumerate}
	\item[(i)] The representation $(H,\StatesPure{H})$ is not necessarily transitive, i.e. it does not necessarily yield a homogeneous space. Pure states belonging to disconnected components cannot be transformed into one another locally, and from an operational perspective they should be thought to correspond to distinct (but incompatible) systems. In quantum theory this will be the case, for example, with pure states belonging to distinct super-selection sectors.

	\item[(ii)] We shall see shortly that whether the composition of two systems can be defined or not depends on the specific connected components of pure states that have been chosen for each system. This is an interesting additional consequence of the decompositional perspective: even assuming that $\orthosubgroups{H}{K}$, in order for two pure states $\Restr{H}{\psi}$ and $\Restr{K}{\varphi}$ to be compatible---i.e. in order for their tensor product to be defined---we need some global state $\gamma$ to exist such that $\Restr{H}{\psi} = \Restr{H}{\gamma}$ and $\Restr{K}{\varphi} = \Restr{K}{\gamma}$.
\end{enumerate}
The considerations above lead finally to the following definition of system in a global process theory.

\begin{definition}
	A \emph{system} in a global process theory $(\Xi,\Theta)$ is a representation in the form $\mathcal{H} = \big(H,H(\Restr{H}{\psi})\big)$, where $H\leq\Xi$ is a self-bicommutant subgroup and $H(\Restr{H}{\psi})$ is the orbit of some pure local state $\Restr{H}{\psi}$ under the action of $H$. We write $\StatesPure{\mathcal{H}} := H(\Restr{H}{\psi})$ for the \emph{pure states} of system $\mathcal{H}$, and $\Transf{\mathcal{H}} := H$ for the \emph{transformations} of system $\mathcal{H}$. We write $I := (\{1\},\{\Xi \psi\})$ for the \emph{trivial system} and $1 := \Xi \psi$ for its only state.
\end{definition}

Note that a system is always a homogeneous space---i.e. it is a transitive representation, by definition---and that the quotient representation $\big(H/Z(H),H(\Restr{H}{\psi})\big)$ itself defines a global process theory, which we refer to as the \emph{restriction of $(\Xi,\Theta)$ to the system $\mathcal{H}$}.

We now proceed to define the tensor product of systems, pending certain compatibility conditions ensuring that any two pure states can always be composed in a unique way. From an operational perspective, the compatibility conditions are a way of ensuring that the compositional version of the purification principle is applied when it comes to tensor products, so that the restriction of a state in $\mathcal{H} \otimes \mathcal{K}$ to the sub-system $\mathcal{H}$ should arise from ``tracing away'' $\mathcal{K}$.

\begin{definition}
	Let $\Xi$ be a group and $H,K \leq \Xi$ be two self-bicommutant subgroups. We say that $H$ and $K$ are \emph{orthocomplementary} if they are orthogonal and they satisfy the following two instances of the orthomodularity law:
	\[
 		K = H' \wedge (H \otimes K)
 		\hspace{3cm}
 		H = K' \wedge (H \otimes K)
	\]
	In particular, $H$ and $H'$ are always orthocomplementary.
\end{definition}

\begin{proposition}
	\label{proposition:self-bicommutants-inside-self-bicommutant}
	Let $\Xi$ be a group, $H,K \leq \Xi$ be two orthogonal self-bicommutant subgroups and $P:=H \vee K$ their join. The commutant of $H$ within $P$ is $H' \wedge P$, and the bicommutant of $H$ within $P$ is $(H' \wedge P)' \wedge P$. If the subgroups $H$ and $K$ are orthocomplementary within $\Xi$ then they are orthogonal, self-bicommutant and orthocomplemented within $P$, in which case within $P$ we have $ H'^{_P} = K$, $K'^{_P} = H$, $H''^{_P} = H$ and $K''^{_P} = K$.
\end{proposition}

When it comes to constructing the tensor product $\mathcal{H} \otimes \mathcal{K}$ of two systems $\mathcal{H}$ and $\mathcal{K}$, orthocomplementarity will mean that the groups of transformations $\Transf{\mathcal{H}}$ and $\Transf{\mathcal{K}}$ will be self-bicommutant not only in the original global process theory $(\Xi,\Theta)$, but also within the restriction of $(\Xi,\Theta)$ to $\mathcal{H} \otimes \mathcal{K}$ (as physical intuition would suggest for sub-systems of a tensor product).

\begin{proposition}
	\label{proposition:orthocomplementary-restriction-equals-trace}
	Let $(\Xi,\Theta)$ be a global process theory, let $H,K \leq \Xi$ be two self-bicommutant subgroups and consider a system $\mathcal{P}$ with $\Transf{\mathcal{P}} = H \otimes K$. If $H$ and $K$ are orthocomplementary, then for any two $\psi,\varphi \in \StatesPure{\mathcal{P}}$ we have that $H'\psi = H' \varphi$ if and only if $K \psi = K \varphi$ (and symmetrically that $K'\psi = K'\varphi$ if and only if $H \psi = H \varphi$).
\end{proposition}

\begin{definition}
	Let $\mathcal{H} = \big(H,H\rho\big)$ and $\mathcal{K} = \big(K,K\sigma\big)$ be two systems in a global process theory $(\Xi,\Theta)$ with $H$ and $K$ orthocomplementary. We say that a state $\psi \in \StatesPure{H \otimes K}$ is a \emph{product state} for $H$ and $K$ if the action of $H \cdot K$ on the orbit $(H \cdot K)\psi$ factors into a product of the action of $H$ on the orbit of $H\Restr{H}{\psi}$ and the action of $K$ on the orbit $K\Restr{K}{\psi}$.
\end{definition}

When the tensor product $\mathcal{H} \otimes \mathcal{K}$ will be defined as a system, the definition of $\psi \in \StatesPure{\mathcal{H} \otimes \mathcal{K}} \subseteq \StatesPure{H \otimes K}$ being a product state will coincide with the definition of $\psi$ being a global product state in the restriction of the global process theory $(\Xi,\Theta)$ to the system $\mathcal{H} \otimes \mathcal{K}$.

\begin{definition}
	Let $\mathcal{H} = \big(H,H\rho\big)$ and $\mathcal{K} = \big(K,K\sigma\big)$ be two systems in a global process theory $(\Xi,\Theta)$. We say that the two systems are \emph{compatible} if the following two conditions hold:
	\begin{enumerate}
		\item $H$ and $K$ are orthocomplementary;
		\item there is some $\psi \in \StatesPure{H \otimes K}$ which is a product over $H$ and $K$ and s.t. $\Restr{H}{\psi} = \rho$ and $\Restr{K}{\psi} = \sigma$.
	\end{enumerate}
	If two systems are compatible, we define their \emph{tensor product} to be the following system:
	\[
		\mathcal{H} \otimes \mathcal{K} := \big(H \otimes K, (H \otimes K)\psi \big)
	\]
\end{definition}
\begin{proposition}
	\label{proposition:tensor-product-well-defined}
	Let $\mathcal{H} = \big(H,H\rho\big)$ and $\mathcal{K} = \big(K,K\sigma\big)$ be two systems in a global process theory $(\Xi,\Theta)$. Compatibility of $\mathcal{H}$ and $\mathcal{K}$ is well-defined, i.e. it does not depend on the choice of representatives $\rho$ and $\sigma$. Given $\rho$ and $\sigma$, the choice of $\psi$ is unique when it exists. Furthermore, the choices of $\psi$ corresponding to different choices of $\rho$ and $\sigma$ are connected under the action of $H \cdot K$. Hence the product $\mathcal{H} \otimes \mathcal{K}$ is also well-defined.
\end{proposition}

From now on, we will switch our focus permanently from self-bicommutant subgroups to systems, and we will adopt the new lighter notation $\Restr{\mathcal{H}}{\rho} := \Restr{\Transf{\mathcal{H}}}{\rho}$ for restriction in the context of systems.

\begin{definition}
	Let $\mathcal{H}$ and $\mathcal{K}$ be two compatible systems in a global process theory $(\Xi,\Theta)$ and consider two pure states $\psi,\varphi \in \StatesPure{\mathcal{H} \otimes \mathcal{K}}$. If $\rho \in \StatesPure{\mathcal{H}}$ and $\sigma \in \StatesPure{\mathcal{K}}$ are pure states in $\mathcal{H}$ and $\mathcal{K}$ respectively, then we define their tensor product as $\rho \otimes \sigma := \psi$, where $\psi$ is the unique pure state in $\StatesPure{\mathcal{H} \otimes \mathcal{K}}$ which is a product over $H$ and $K$ and such that  $\Restr{\mathcal{H}}{\psi} = \rho$ and $\Restr{\mathcal{K}}{\psi} = \sigma$.
\end{definition}

The tensor product of systems was so far defined in a binary way, but the following result shows that it can be extended to arbitrary collections by iteration, in a strictly commutative and associative way.
\begin{proposition}
	\label{proposition:tensor-product-associative}
	Let $\mathcal{H},\mathcal{K},\mathcal{L}$ be systems in a global process theory $(\Xi,\Theta)$. Let $\mathcal{H}$ be compatible with $\mathcal{K}$, so that the tensor product $\mathcal{H} \otimes \mathcal{K}$ can be defined, and let $\mathcal{H} \otimes \mathcal{K}$ be in turn compatible with $\mathcal{L}$, so that the tensor product $(\mathcal{H} \otimes \mathcal{K}) \otimes \mathcal{L}$ can be defined. Then $\mathcal{H}$ and $\mathcal{K}$ are both compatible with $\mathcal{L}$, so that the tensor products $\mathcal{H} \otimes \mathcal{L}$ and $\mathcal{K} \otimes \mathcal{L}$ can be defined. Furthermore $\mathcal{H}$ is compatible with $\mathcal{K} \otimes \mathcal{L}$ and $\mathcal{K}$ is compatible with $\mathcal{H} \otimes \mathcal{L}$, so that the tensor products $\mathcal{H} \otimes (\mathcal{K} \otimes \mathcal{L})$ and $\mathcal{K} \otimes (\mathcal{H} \otimes \mathcal{L})$ can be defined. All tensor products are strictly commutative and strictly associative.
\end{proposition}

\section{States and Processes}
\label{section:systems=and-processes}

We now have all the pieces necessary to define a symmetric partially-monoidal category of systems with pure states and ``pure processes'' between them. In so many words, a symmetric partially-monoidal category is similar to a symmetric monoidal category, but where the tensor product between objects $A \otimes B$ may not always be defined. When the tensor products $A \otimes B$ and $C \otimes D$ are defined, however, so is the tensor product $f \otimes g$ of all morphisms $f: A \rightarrow C$ and $g:B \rightarrow D$. For a rigorous definition of symmetric partially-monoidal categories, refer to Appendix \ref{appendix:partially-monoidal-categories}.
\begin{theorem}
	\label{theorem:pure-process-theory}
	Let $(\Xi,\Theta)$ be a global process theory. Then the following defines a symmetric partially-monoidal category $\ProcessTheoryPure{\Xi}{\Theta}$, which we refer to as the \emph{pure process theory arising from $(\Xi,\Theta)$}:
	\begin{enumerate}
		\item Objects are all systems $\mathcal{H}$, equipped with their tensor product and the trivial system $I$ as tensor unit.
		\item Whenever $\mathcal{H}$ and $\mathcal{L}$ are compatible, morphisms $\mathcal{H} \rightarrow \mathcal{H} \otimes \mathcal{L}$ are defined to be the following injective maps $\StatesPure{\mathcal{H}} \rightarrow \StatesPure{\mathcal{H} \otimes \mathcal{L}}$:
		\[
			(u,\sigma) := \rho \mapsto u(\rho \otimes \sigma)
		\]
		for some choice of pure state $\sigma \in \StatesPure{\mathcal{L}}$ and transformation $u \in \Transf{\mathcal{H}\otimes \mathcal{L}}$.
		We refer to the morphisms in $\ProcessTheoryPure{\Xi}{\Theta}$ as \emph{pure processes}.
		\item If $(u,\sigma): \mathcal{H} \rightarrow \mathcal{H}\otimes \mathcal{K}$ and $(v,\tau): \mathcal{H} \otimes \mathcal{K} \rightarrow \mathcal{H}\otimes \mathcal{K} \otimes \mathcal{L}$ are pure processes, then their composition is defined to be:
		\[
			(v,\tau) \circ (u,\sigma) := (vu,\sigma \otimes \tau): \mathcal{H} \rightarrow (\mathcal{H}\otimes \mathcal{K}) \otimes \mathcal{L}
		\]

		\item If $(u,\sigma): \mathcal{H} \rightarrow \mathcal{H}\otimes \mathcal{K}$ and $(v,\tau): \mathcal{L} \rightarrow \mathcal{L}\otimes \mathcal{M}$ are pure processes with compatible domains and codomains, then their tensor product is defined to be:
		\[
			(u,\sigma) \otimes (v,\tau) := (u \otimes v, \sigma \otimes \tau): \mathcal{H} \otimes \mathcal{L} \longrightarrow (\mathcal{H}\otimes \mathcal{K}) \otimes (\mathcal{L}\otimes \mathcal{M})
		\]
	\end{enumerate}
\end{theorem}

Having defined a process theory with systems and pure processes, we are now almost done: all that stands between us and a full process theory is the introduction of a discarding map. Contrary to the traditional compositional picture, however, discarding a sub-system does not mean that we have simply forgotten it: in order for things to be well-defined, we must make sure never to use that sub-system again. This means that the objects of the full process theory will be \emph{pairs} of systems: the system which we wish to act upon and the environment discarded so far. In particular, the non-pure states available on a system depend on the environment that was discarded during their creation.
\begin{definition}
	A \emph{system-environment pair} is a pair $(\mathcal{H},\mathcal{E})$ of compatible systems: we refer to $\mathcal{H}$ as the \emph{system} and to $\mathcal{E}$ as the \emph{environment}. The \emph{states} on a system-environment pair $(\mathcal{H},\mathcal{E})$ are given by the following space:
	\[
		\States{(\mathcal{H},\mathcal{E})} := \suchthat{\Restr{\mathcal{H}}{\psi}}{\psi \in \StatesPure{\mathcal{H}\otimes\mathcal{E}}}
	\]
\end{definition}

Note that, independently of $\mathcal{E}$, we always have the following inclusions of $\Transf{\mathcal{H}}$-representations:
\[
	\StatesPure{\mathcal{H}} = \States{(\mathcal{H},I)} \subseteq \States{(\mathcal{H},\mathcal{E})} \subseteq \States{\Transf{\mathcal{H}}}
\]
This means that can identify a system-environment pair $(\mathcal{H},\mathcal{E})$ with a representation of $\Transf{\mathcal{H}}$---namely that defined by $h(\Restr{\mathcal{H}}{\psi}):= \Restr{\mathcal{H}}{h\psi}$ for all $h \in \Transf{\mathcal{H}}$---having the homogeneous space $\mathcal{H}$ as a sub-representation.

If $(\mathcal{H},\mathcal{E})$ is a system-environment pair and $\mathcal{H} = \mathcal{K} \otimes \mathcal{F}$, then iterated restriction $\rho \mapsto \Restr{K}{\rho}$ gives a well-defined map $\States{(\mathcal{H},\mathcal{E})} \rightarrow \States{(\mathcal{K},\mathcal{E} \otimes \mathcal{F})}$, which is an intertwiner for the action of $\mathcal{K}$ on both spaces. In the special case where $\mathcal{K} = I$, iterated restriction yields the unique map $\top:\States{(\mathcal{H},\mathcal{E})} \rightarrow \States{(I,\mathcal{E} \otimes \mathcal{H})}$, sending every $\rho \in \States{(\mathcal{H},\mathcal{E})}$ to $1$, the unique state in $\States{(I,\mathcal{E}\otimes \mathcal{H})}$.
\begin{definition}
	Given two system-environment pairs $(\mathcal{H},\mathcal{E})$ and $(\mathcal{K},\mathcal{F})$, we say that they are \emph{compatible} if $\mathcal{H} \otimes \mathcal{E}$ is compatible with $\mathcal{K} \otimes \mathcal{F}$ (which in turn implies that $\mathcal{H}$ is compatible with $\mathcal{K}$ and that $\mathcal{E}$ is compatible with $\mathcal{F}$):
	The \emph{tensor product} of two compatible system-environment pairs $(\mathcal{H},\mathcal{E})$ and $(\mathcal{K},\mathcal{F})$ is defined to be:
	\[
		(\mathcal{H},\mathcal{E}) \otimes (\mathcal{K},\mathcal{F}) := (\mathcal{H}\otimes\mathcal{H},\mathcal{E}\otimes\mathcal{F})
	\]
	The tensor product is strictly associative and commutative, and has the \emph{trivial system-environment pair} $(I,I)$ as bilateral unit.
\end{definition}

\begin{theorem}
	\label{theorem:process-theory}
	Let $(\Xi,\Theta)$ be a global process theory. Then the following defines a symmetric partially-monoidal category $\ProcessTheory{\Xi}{\Theta}$, which we refer to as the \emph{process theory arising from $(\Xi,\Theta)$}:
	\begin{enumerate}
		\item Objects are system-environment pairs $(\mathcal{H},\mathcal{E})$, equipped with their tensor product and the trivial system-environment pair $(I,I)$ as their tensor unit.
		\item Morphisms only exist with types $(\mathcal{H},\mathcal{E}) \rightarrow (\mathcal{K},\mathcal{E}\otimes\mathcal{M})$, in which case they are the maps $(u,\sigma;\mathcal{M}):\States{(\mathcal{H},\mathcal{E})} \rightarrow \States{(\mathcal{K}, \mathcal{E} \otimes \mathcal{M})}$ taking the following form:
		\[
			(u,\sigma;\mathcal{M})
			:=
			\rho
			\mapsto
			\Restr{\mathcal{K}}{u(\rho \otimes \sigma)}
		\]
		where $(\mathcal{L},I)$ is compatible with $(\mathcal{H},\mathcal{E})$, $\sigma$ is a pure state in $\StatesPure{\mathcal{L}}$, we have the equality of tensor products $\mathcal{H} \otimes \mathcal{L} = \mathcal{K} \otimes \mathcal{M}$ and $u$ is a transformation in $\Transf{\mathcal{H} \otimes \mathcal{L}}$. Note that  $(I,\mathcal{M})$ is always necessarily compatible with $(\mathcal{K},\mathcal{E})$. We refer to the morphisms in $\ProcessTheory{\Xi}{\Theta}$ as \emph{processes}.

		\item If $(u,\sigma;\mathcal{B}): (\mathcal{H},\mathcal{A}) \rightarrow (\mathcal{K}, \mathcal{A}\otimes \mathcal{B})$ and $(v,\tau;\mathcal{C}): (\mathcal{K},\mathcal{A} \otimes \mathcal{B}) \rightarrow (\mathcal{L}, (\mathcal{A}\otimes \mathcal{B}) \otimes \mathcal{C})$ are processes, then their composition is defined to be:
		\[
			(v,\tau;\mathcal{C}) \circ (u,\sigma;\mathcal{B}) := (vu,\sigma \otimes \tau; \mathcal{B} \otimes \mathcal{C}): (\mathcal{H},\mathcal{A}) \rightarrow (\mathcal{L}, \mathcal{A}\otimes (\mathcal{B} \otimes \mathcal{C}))
		\]

		\item If $(u,\sigma;\mathcal{B}): (\mathcal{H},\mathcal{A}) \rightarrow (\mathcal{K}, \mathcal{A}\otimes \mathcal{B})$ and $(v,\tau;\mathcal{D}): (\mathcal{L},\mathcal{C}) \rightarrow (\mathcal{M}, \mathcal{C}\otimes \mathcal{D})$ are processes with compatible domains and compatible codomains, then their tensor product is defined to be:
		\[\hspace{-5mm}
			(u,\sigma;\mathcal{B}) \otimes (v,\tau;\mathcal{D}) := (u \otimes v, \sigma \otimes \tau;\mathcal{B}\otimes \mathcal{D}): (\mathcal{H} \otimes \mathcal{L}, \mathcal{A} \otimes \mathcal{C}) \longrightarrow (\mathcal{K} \otimes \mathcal{M}, (\mathcal{A}\otimes \mathcal{C}) \otimes (\mathcal{B} \otimes \mathcal{D}))
		\]
	\end{enumerate}
\end{theorem}

We can think of $\ProcessTheory{\Xi}{\Theta}$ as the theory that one obtains from $\ProcessTheoryPure{\Xi}{\Theta}$ by introducing the \emph{discarding maps}, the unique process $\top: (\mathcal{H},\mathcal{E}) \rightarrow (I,\mathcal{H} \otimes \mathcal{E})$. The following result makes this intuition rigorous and implies that the (compositional) purification principle is always satisfied by $\ProcessTheory{\Xi}{\Theta}$.
\begin{proposition}
	\label{proposition:purification-principle}
	Let $(\Xi,\Theta)$ be a global process theory. The pure process theory $\ProcessTheoryPure{\Xi}{\Theta}$ is isomorphic to the symmetric partially-monoidal sub-category of the process theory $\ProcessTheory{\Xi}{\Theta}$ given by the objects in the form $(\mathcal{H},I)$---i.e. those with trivial environment---and morphisms in the form $(u,\sigma;I)$---i.e. those with trivial discarding. The smallest symmetric partially-monoidal sub-category of $\ProcessTheory{\Xi}{\Theta}$ which contains all transformations $(u,1;I):(\mathcal{H},I) \rightarrow (\mathcal{H},I)$ and all pure states $(1,\sigma;I):(I,I) \rightarrow (\mathcal{H},I)$ is the pure process theory $\ProcessTheoryPure{\Xi}{\Theta}$. The smallest symmetric partially-monoidal sub-category of $\ProcessTheory{\Xi}{\Theta}$ which contains the pure process theory $\ProcessTheoryPure{\Xi}{\Theta}$ and all discarding maps is $\ProcessTheory{\Xi}{\Theta}$ itself.
\end{proposition}

The discarding maps are the only ``effects'' in the theory, as long as we are willing to broaden our definition of effects to include all morphisms with codomain in the form $(I,\mathcal{E})$--the trivial system, together with some environment. Under this broader definition, the process theory $\ProcessTheory{\Xi}{\Theta}$ is causal.
\begin{proposition}
	\label{proposition:causality}
	Let $(\Xi,\Theta)$ be a global process theory, and say that a map in $\ProcessTheory{\Xi}{\Theta}$ is a \emph{generalised effect} if its codomain is in the form $(I,\mathcal{E})$. Then $\ProcessTheory{\Xi}{\Theta}$ is causal in the sense that every system-environment pair $(\mathcal{H},\mathcal{E})$ has a unique generalised effect, the discarding map $\top: (\mathcal{H},\mathcal{E}) \rightarrow (I,\mathcal{E} \otimes \mathcal{H})$.
\end{proposition}

\section{Quantum Theory}
\label{section:quantum-theory}

In quantum theory we only care about the type of a quantum system, not its physical realisation. To recover quantum theory, we begin by considering the global process theory given by an $n$-dimensional projective Hilbert space $\Theta_n$ with the action of the centreless group $\Xi_n := U(n)/U(1)$ on it. If $H$ is a subgroup, then one can consider the projective unitary representation of $H$ on $\Theta_n$ induced by the injection $H \hookrightarrow U(n)/U(1)$.
Standard results from representation theory (see e.g. \cite{chiribella2018agents}) can be used to deduce that self-bicommutant groups $H$ and their respective commutants $H'$ are those with elements ($u \in H$ on the left, $v \in H'$ on the right) taking the following form modulo global phase:
\[
	u = \bigoplus_{j=1}^{s_H} U_j \otimes 1_{B_j^{(H)}}
	\hspace{4cm}
	v = \bigoplus_{j=1}^{s_H} 1_{A_j^{(H)}} \otimes V_j
\]
where $\bigoplus_{j=1}^{s_H} A_j^{(H)} \otimes B_j^{(H)}$ is some fixed direct sum decomposition of $\Theta_n$, $U_j$ are arbitrary unitary transformations on $A_j^{(H)}$ and $V_j$ are arbitrary unitary transformations on $B_j^{(H)}$.

The subspaces $\big(A_j^{(H)} \otimes B_j^{(H)}\big)_{j=1}^{s_H}$ associated to $H$ are global super-selection sectors and there are exactly $s_H$ distinct (and mutually incompatible) systems having $H$ as their group of local transformations, namely the projective Hilbert spaces $\big(A_j^{(H)}\big)_{j=1}^{s_H}$ with the action of $H$ on them. The centre of $H$ contains the possible relative phases across the super-selection sectors, i.e. we have $Z(H) = \big(\oplus_{j=1}^{s_H} U(1)\big) / U(1) \cong U(1)^{s_H - 1}$.

Two special cases are of particular interest, because their combination generates everything else: the \emph{purely additive} and the \emph{purely multiplicative} sub-systems of the global system $(\Xi_n,\Theta_n)$.

By \emph{purely additive} sub-systems of $(\Xi_n,\Theta_n)$ we mean those corresponding to self-bicommutant subgroups $H$ with direct sum decomposition $(A \otimes \mathbb{C}) \oplus (\mathbb{C} \otimes B)$, where the only non-trivial system with $H$ as its group of transformations is $\mathcal{A}:=\big((U(A)/U(1)) \oplus 1_B,A\big)$. Similarly, $\mathcal{B}:=\big(1_A \oplus (U(B)/U(1)),B\big)$ is the only non-trivial system with $H'$ as its group of transformations. Because these systems arise by super-selection, physical intuition would mandate that they be incompatible: this is indeed the case since $H$ and $H'$ are orthogonal but not orthocomplementary, with $H \otimes H' = (U(A) \oplus U(B))/U(1) \neq \Xi_n$.

By \emph{purely multiplicative} sub-systems of $(\Xi_n,\Theta_n)$ we mean those corresponding to self-bicommutant subgroups $H$ with direct sum decomposition $A \otimes B$, i.e. those with a single super-selection sector ($s_H = 1$). There is a single system with $H$ as its group of transformations, namely $\mathcal{A}:=\big((U(A)/U(1)) \otimes 1_B,A\big)$, and similarly there is a single system with $H'$ as its group of transformations, namely $\mathcal{B}:=\big(1_A \otimes (U(B)/U(1)),B\big)$. The systems $\mathcal{A}$ and $\mathcal{B}$ are always compatible and we have $\mathcal{A} \otimes \mathcal{B} = (\Xi_n,\Theta_n)$.

From this description we see that all projective Hilbert spaces with dimension up to $n$---along with all their pure states and all their unitary automorphisms---arise in $\ProcessTheory{\Xi_n}{\Theta_n}$. However, all mixed states arise only for Hilbert spaces of dimension up to $\lfloor\sqrt{n}\rfloor$. Worse still, some CPTP maps arise, but their sequential and parallel composition are limited by the (eventual) exhaustion of compatible systems available (in the case of sequential composition, to use for the ancillary states and for discarding).

This limitation of $\ProcessTheory{\Xi_n}{\Theta_n}$ should not, by this point, come as any surprise: we have willingly restricted ourselves to a theory with a finite-dimensional global Hilbert space, so we are bound to run out of space at some point. If we want to recover quantum theory, we must now do two things:
\begin{enumerate}
	\item Take a limit in which $n \rightarrow \infty$, so we have an unlimited supply of compatible systems to choose from whenever we want a quantum system of a given type.
	\item Forget the individual identities of systems, because we only care about their type---in quantum theory we want ``a qubit'', not ``that qubit over there''.
\end{enumerate}
Because our definitions are purely group-theoretic, the first part is easy. For the group of global transformations we consider the following limit:
\[
	\Xi:= \lim_{n \rightarrow \infty} U(n)/U(1) = \bigcup_{n=2}^{\infty} U(n)/U(1)
\]
where the nesting $U(n)/U(1) \hookrightarrow U(n+1)/U(1)$ is the one corresponding to the injection of Hilbert spaces $\Theta_n \hookrightarrow \Theta_{n+1} := \Theta_n \oplus \mathbb{C}$. For the global space, consider the corresponding $\ell_0$ space $\Theta := \mathbb{C}^{\mathbb{N}} := \lim_{n\rightarrow \infty} \Theta_n$. Amongst the systems of this ``limit'' global process theory $(\Xi,\Theta)$ there now are infinitely many compatible ``copies'' of projective Hilbert spaces for all finite dimensions.

The second part is also easy: whenever taking the tensor product of two finite-dimensional Hilbert spaces, we simply pick two systems with the appropriate types which are compatible. In the limit $n \rightarrow \infty$, there is always going to be an endless supply of compatible systems available, yielding arbitrary finite tensor products of finite-dimensional Hilbert spaces. This means that all states and CPTP maps will be defined and freely composable both in sequence and in parallel. We have recovered quantum theory.

A rigorous formalisation of this procedure and a generalisation to other examples of ``limit'' global process theories will be covered in future versions of this work.

\section{Acknowledgements}

The author's heartfelt thanks go to David Reutter: without his many categorical and algebraic insights, this work would look very different than it does today. This publication was made possible through the support of a grant from the John Templeton Foundation. The opinions expressed in this publication are those of the authors and do not necessarily reflect the views of the John Templeton Foundation.

\bibliographystyle{eptcs}
\bibliography{biblio}

\appendix

\newcommand{\CatCategory}{\operatorname{Cat}}
\newcommand{\PCatCategory}{\operatorname{PCat}}
\newcommand{\supp}[1]{\operatorname{supp}\left(#1\right)}
\newcommand{\obj}[1]{\operatorname{obj}\left(#1\right)}

\section{Partially-monoidal categories}
\label{appendix:partially-monoidal-categories}

\begin{definition}
	We define the monoidal $(2,1)$-category $\PCatCategory$ of \emph{categories and partial functors} to be given by the following data:
	\begin{enumerate}
		\item The objects of $\PCatCategory$ are categories.
		\item The morphisms $\mathcal{C} \rightharpoonup \mathcal{D}$ in $\PCatCategory$, known as the \emph{partial functors}, are spans $(I,F):\mathcal{C} \hookleftarrow \mathcal{X} \rightarrow \mathcal{D}$ of functors, where $I:\mathcal{X} \hookrightarrow \mathcal{C}$ defines a replete full subcategory of $\mathcal{C}$ and $F: \mathcal{X} \rightarrow \mathcal{D}$ is any functor. We say that the subcategory defined by $I$ is the \emph{support} of $(I,F)$ and write $\supp{(I,F)}$ for it.
		\item Composition of morphisms is by pullback of spans:
		\[
			(J,G) \circ (I,F) := \big(I \circ F^{*}(J),G  \circ J^{*}(F)\big)
		\]
		where the pullback $F^{*}(J)$ of $J$ along $F$ defines the (necessarily replete) full subcategory $\mathcal{Z}$ of $\mathcal{X}$ spanned by the objects $A \in \obj{\mathcal{X}}$ such that $F(A) \in \supp{(J,G)}$, while the pullback $J^{*}(F)$ of $F$ along $J$ is the restriction of $F$ to $\mathcal{Z}$.
		\item Given two parallel morphisms $(I,F):\mathcal{C} \hookleftarrow \mathcal{X} \rightarrow \mathcal{D}$ and $(J,G):\mathcal{C} \hookleftarrow \mathcal{Y} \rightarrow \mathcal{D}$, the 2-morphisms $(I,F) \Rightarrow (J,G)$ are pairs $(\Phi,\alpha)$ of an isomorphism of categories $\Phi: \mathcal{X} \rightarrow \mathcal{Y}$ such that $J \circ \phi = I$ and a natural isomorphism $\alpha: F \rightarrow G \circ \Phi$. In particular, 2-cells $(I,F) \Rightarrow (J,G)$ exist only when the supports $\supp{(I,F)}$ and $\supp{(J,G)}$ coincide. Vertical and horizontal composition of 2-morphisms are defined in the obvious way by diagram-chasing.
		\item The tensor product on objects is the Cartesian product of categories $\mathcal{C} \times \mathcal{D}$. The tensor product $(I,F) \times (J,G)$ on morphisms is $(I \times J, F \times G)$. As is the case for the tensor product in the category of sets and partial functions, note that the tensor product on $\PCatCategory$ is not Cartesian, despite the notation.
	\end{enumerate}
\end{definition}

The definition above is closely aligned to the definition of partial functors from \cite{coecke2013causal}, the two main differences being the introduction of 2-morphisms and the requirement that the support be a full and replete sub-category (rather than a generic subcategory).

\begin{definition}
	A \emph{partially-monoidal category} is a weak pseudomonoid in the monoidal $(2,1)$-category $\PCatCategory$ of categories and partial functors. A partially-monoidal category is \emph{strict} when the weak pseudomonoid is a monoid. A partially-monoidal category is \emph{symmetric} when the weak pseudomonoid is weakly commutative.
\end{definition}

The requirement that the support of partial functors be a full subcategory means that whenever the tensor products $A \otimes C$ and $B \otimes D$ in a partially-monoidal category $\mathcal{C}$ are defined then so is the tensor product $f \otimes g$ for all pairs of morphisms $f: A \rightarrow B$ and $g: C \rightarrow D$. The requirement that the support of partial functors be a replete subcategory means that whenever the tensor product $A \otimes B$ in a partially-monoidal category then so is $C \otimes D$ for all $C \cong A$ and all $D \cong B$. The definition of 2-morphisms in $\PCatCategory$ means that $A \otimes (B \otimes C)$ is defined if and only if $(A \otimes B) \otimes C$ is defined\footnote{Because the existence of a 2-isomorphism for associativity implies that the support for the partial functors $\otimes \circ (1_{\mathcal{C}} \times \otimes)$ and $\otimes \circ (\otimes \times 1_{\mathcal{C}})$ is the same replete full subcategory of $\mathcal{C} \otimes \mathcal{C} \otimes \mathcal{C}$.}, in which case $A \otimes (B \otimes C) \cong (A \otimes B) \otimes C$.

\section{Proofs}

\subsubsection*{Proposition \ref{proposition:tensor-product-transformations}}
\begin{proof}
	Properties of $\otimes$ on subgroups all follows from its definition as the $\vee$ operation. Associativity on transformations follows from associativity of group multiplication in $\Xi$. Finally, the exchange law is a consequence of the fact that $H$ and $K$ are orthogonal:
	\[
		c \in K, \, b \in H
		\hspace{3mm}\Rightarrow\hspace{3mm}
		bc = cb
		\hspace{3mm}\Rightarrow\hspace{3mm}
		(ab) \otimes (cd) = abcd = acbd = (a \otimes c)(b \otimes d)
	\]
	Because $hk = kh$ for all $h \in H$ and $k \in K$, it is trivial to see that the identity map $H \otimes K \rightarrow K \otimes H$ behaves as described.
\end{proof}

\subsubsection*{Proposition \ref{proposition:systems-action}}
\begin{proof}
	Because every $h \in H$ commutes with the entirety of $H'$, we immediately have that:
	\[
		h(\Restr{H}{\psi}) = hH'\psi = H'h\psi = \Restr{H}{h\psi}
	\]
	If $h \in Z(H)$, then we also have that $h \in H'$, so that:
	\[
		h(\Restr{H}{\psi}) = hH'\psi = H'\psi = \Restr{H}{\psi}
	\]
	Because the centre $Z(H)$ acts trivially, we have the following representation of the quotient $H/Z(H)$:
	\[
		(Z(H)h)\big(\Restr{H}{\psi}\big) = Z(H)hH'\psi = h Z(H) H' \psi = h H' \psi = h(\Restr{H}{\psi})
	\]
	This concludes our proof.
\end{proof}

\subsubsection*{Proposition \ref{proposition:iterated-restriction-well-defined}}
\begin{proof}
	Recall that $\Restr{H}{\psi} = H'\psi$ and $\Restr{H}{\varphi} = H'\varphi$, so that the condition on $\psi,\varphi$ can be rewritten as $H'\psi = H'\varphi$. If $K \leq H$ then $H' \leq K'$, and hence:
	\[
		\Restr{K}{\psi} = K'\psi = K'H'\psi = K'H'\varphi = K'\varphi = \Restr{K}{\varphi}
	\]
	If $H = \Xi$, then $H' = \{1\}$ and $\Restr{H}{\psi} = \{\psi\}$ can be canonically identified with $\psi$. Also:
	\[
		\Restr{K}{\Restr{H}{\psi}} = K'\{\psi\} = K'\psi = \Restr{K}{\psi}
	\]
	This concludes our proof.
\end{proof}

\subsubsection*{Proposition \ref{proposition:pure-local-states-representation}}
\begin{proof}
	Let $\Restr{H}{\psi}$ be pure, and consider $h(\Restr{H}{\psi}) := \Restr{H}{h\psi}$. Because we are in the same $H$-orbit, all we need to do is compute the three stabilisers for $h\psi$ and check that they factor. On the left hand side, we have:
	\[
		\Stab{\Theta}{H\cdot H'}{h\psi}  = h \Stab{\Theta}{H\cdot H'}{\psi} h^{-1}
	\]
	Elements in the subgroup above take the form $h ab h^{-1}$ for all $a \in H$ and $b \in H'$ such that $ab\psi = \psi$. Because $\psi$ is a product over $H$ (eq'tly over $H'$), this is the same as taking elements in the form $h ab h^{-1}$ for all $a \in H$ such that $a \psi = \psi$ and all $b \in H'$ such that $b \psi = \psi$. On the right hand side, we have:
	\[
		\Stab{\Theta}{H}{h\psi} \cdot \Stab{\Theta}{H'}{\psi}
		=
		\left(h \Stab{\Theta}{H}{\psi} h^{-1}\right) \cdot \left(h \Stab{\Theta}{H'}{\psi} h^{-1}\right)
	\]
	Elements in the subgroup above take the following form $h a h^{-1} h b h^{-1} = h ab h^{-1}$ for all $a \in H$ such that $a\psi = \psi$ and all $b \in H'$ such that $b \psi = \psi$. We conclude that $h(\Restr{H}{\psi}) := \Restr{H}{h\psi}$ is also pure. The stabiliser for $\Restr{H}{h\psi}$ in $H$ takes the following form:
	\[
		\Stab{\StatesPure{H}}{H}{\Restr{H}{\psi}}
		=
		\suchthat{h \in H}{h H' \psi = H' \psi}
		=
		\suchthat{h \in H}{\Forall{a \in H'}{\Exists{b \in H'}{ h a \psi = b \psi}}}
	\]
	The last equation can be rewritten as $h (b^{-1}a) \psi = \psi$, using the fact that $b^{-1}h = hb^{-1}$, and the requirement that $\psi$ be a product over $H$ further implies that $h \psi = \psi$:
	\[
		\Stab{\StatesPure{H}}{H}{\Restr{H}{\psi}}
		=
		\suchthat{h \in H}{h \psi = \psi}
		=
		\Stab{\Theta}{\Xi}{\psi} \cap H
	\]
	This concludes our proof.
\end{proof}

\subsubsection*{Corollary \ref{corollary:states-with-pure-marginals-are-product}}
\begin{proof}
	Because the definition of being a product is symmetrical with respect to taking the commutant, we know that $\psi$ is a product over $H$ if and only if it is a product over $H'$. If $H'\psi = H'\varphi$, then $\varphi = k \psi$ for some $k \in H'$. If $\psi$ is a product over $H$, then it is a product over $H'$: by the previous result, we conclude that $\varphi$ is a product over $H'$ as well, and hence also a product over $H$.
\end{proof}

\subsubsection*{Proposition \ref{proposition:self-bicommutants-inside-self-bicommutant}}
\begin{proof}
	The commutant of $H$ within $P:= H \vee K$ is clearly $H'^{_P} = H' \wedge P$, and the commutant of $K$ within $P$ is $K'^{_P} = K' \wedge P$. The bicommutant of $H$ within $P$ is then clearly $H''^{_P} = (H' \wedge P)' \wedge P$, and the bicommutant of $K$ within $P$ is $K''^{_P} = (K' \wedge P)' \wedge P$. We immediately deduce the following identity:
	\[
		((H' \wedge P) \vee P') \wedge P = ((H' \wedge P)' \wedge P)' \wedge P = H'''^{_P} = H'^{_P} = (H' \cap P)
	\]
	If $H$ and $K$ are orthocomplementary, then within $P$ we immediately have $H'^{_P} = (H' \wedge P) = K$ and $K'^{_P} = (K' \wedge P) = H$ so that $H$ and $K$ are both orthogonal and self-bicommutant within $P$, with $H$ the orthocomplement of $K$ within $P$ and vice versa.
\end{proof}

\subsubsection*{Proposition \ref{proposition:orthocomplementary-restriction-equals-trace}}
\begin{proof}
	If $H$ and $K$ are orthocomplementary, then in particular they are orthogonal and we have $H \leq K'$ and $K \leq H'$: hence $K \psi = K \varphi$ always implies $H'\psi = H'\varphi$, and $H \psi = H \varphi$ always implies $K'\psi = K'\varphi$. Now assume that $H' \psi = H' \varphi$, so that $\psi = u \varphi$ for some $u \in H' \wedge (H \otimes K)$ (because the action of $H \otimes K$ is transitive on the homogeneous space $\mathcal{P}$). Because of orthocomplementarity, we have that $H' \wedge (H \otimes K) = K$, so that $u \in K$ and $K \psi = K \varphi$. Symmetrically, if $K'\psi = K'\varphi$ then $H \psi = H \varphi$.
\end{proof}

\subsubsection*{Proposition \ref{proposition:tensor-product-well-defined}}
\begin{proof}
	If $h\rho$ and $k\sigma$ are other representatives, then the state $hk \psi \in \StatesPure{H \otimes K}$ exists and satisfies $\Restr{H}{hk\psi} = h\Restr{H}{\psi} = h\rho$ and $\Restr{K}{hk\psi} = k\Restr{K}{\psi} = k\sigma$; hence compatibility is independent of the representatives $\rho$ and $\sigma$. Now let $\psi,\varphi \in \StatesPure{H \otimes K}$ be product states over $H$ and $K$ such that $H'\psi = \rho = H' \varphi$ and $K'\psi = \sigma = K' \varphi$. From $H' \psi = H' \varphi$ and orthocomplementarity we deduce that $\psi = k \varphi$ for some $k \in K$, and symmetrically from $K' \psi = K' \varphi$ we deduce that $\psi = h \varphi$ for some $h \in H$. Putting the two equations together we get $hk \varphi = \varphi$, and from the fact that $\varphi$ is a product state over $H$ and $K$ we deduce that $h \varphi = \varphi$ and $k\varphi = \varphi$, so that $\psi = \varphi$.
\end{proof}

\subsubsection*{Proposition \ref{proposition:tensor-product-associative}}
\begin{proof}
	Write $\mathcal{H} = (H,\rho)$, $\mathcal{K} = (K,\sigma)$, $\mathcal{L} = (L,\tau)$, $\mathcal{H} \otimes \mathcal{K} = (H \otimes K, \psi)$ and $(\mathcal{H} \otimes \mathcal{K}) \otimes \mathcal{L} = (H\otimes K \otimes L, \varphi)$. If $\mathcal{H} \otimes \mathcal{K}$ is compatible with $\mathcal{L}$, then $H \otimes K$ is orthocomplementary to $L$. From $H \otimes K \leq L'$ we deduce $H,K \leq L'$, so that $\orthosubgroups{H}{L}$ and $\orthosubgroups{K}{L}$.

	We now write $P:=H \otimes K$ for short, and without loss of generality we work within the group $Q := H \otimes K \otimes L$: except when explicitly state otherwise, all commutants are taken within $Q$. Because $P$ and $L$ are orthocomplementary within $\Xi$, they are both self-bicommutant (within $Q$) and we have $P' = L$ and $L' = P$ (within $Q$). Because $H$ and $K$ are orthocomplementary within $\Xi$ we have $K = H'^{_\Xi} \wedge P = (H'^{_\Xi} \wedge Q) \wedge P = H' \wedge P$, and symmetrically $H = (K' \wedge P)$, so that $H$ and $K$ are also self-bicommutant and orthocomplementary within $Q$. We have $(H \otimes L)' = (H \vee P')' = H' \wedge P = K$ and similarly that $(K \otimes L)' = H$, $H \otimes L = K'$ and $K \otimes L = H'$. From those equations we deduce that $H' \wedge (H \otimes L) = H' \wedge K' = (H \vee K)' = P' = L$ and similarly that $K'\wedge (K \otimes L) = L$. Finally, $L' \wedge (H \otimes L) = (H \vee P') \wedge P = ((H' \wedge P) \wedge P)' =  (K \wedge P)' = H$ and similarly $L' \wedge (K \otimes L) = K$. Hence $H$ and $L$ are orthocomplementary, and so are $K$ and $L$, $H\otimes L$ and $K$ and $K \otimes L$ and H.

	Finally, $\varphi$ being a product state implies that the action of $(H\otimes K) \cdot L$ on $\varphi$ factors into the product of the action of $H \otimes K$ on $\psi$ and of $L$ on $\tau$. Then action of $H \cdot K \cdot L$ on $\varphi$ hence factors into the product of the action of $H \cdot K$ on $\psi$ and the action of $L$ on $\tau$; in turn, the action of $H \cdot K$ on $\psi$ factors into the product of the action of $H$ on $\rho$ and the action of $K$ on $\sigma$. All in all, the action of $H \cdot K \cdot L$ on $\varphi$ factors into the product of the actions of $H$ on $\rho$, $K$ on $\sigma$ and $L$ on $\tau$. Hence, the action of $K \cdot L$ on $\Restr{K \otimes L}{\varphi}$ factors into the product of the action of $K$ on $\sigma$ and of $L$ on $\tau$, and similarly the action of $H \cdot L$ on $\Restr{H \otimes L}{\varphi}$ factors into the product of the action of $H$ on $\sigma$ and of $L$ on $\tau$. This means that all products listed in the statement of this proposition can be defined by considering $\varphi$ and its restrictions as the relevant product states. In particular, all tensor products are strictly commutative and associative.
\end{proof}

\subsubsection*{Theorems \ref{theorem:pure-process-theory} and \ref{theorem:process-theory}}
\begin{proof}
	The essential points of both proofs are the same, so we will group them into a single proof: the nitty-gritty details that differ between the two are all self-evident and straightforward applications of previous results, so they are omitted here. The main ingredient in the proof is Proposition \ref{proposition:tensor-product-associative}, which finds dual use:
	\begin{enumerate}
		\item It allows all required compatibility conditions between individual systems to be expressed compactly as compatibility conditions between tensor products.
		\item It ensures that both categories are equipped with an operation $\otimes$ which satisfies the conditions for the tensor product of a strict symmetric partially-monoidal category.
	\end{enumerate}
	The reason why composition and tensor product are well-defined become immediately apparent if we adopt a diagrammatic representation for processes (pure processes on the left, general processes on the right---we remark the ``trivial'' role of objects in the form $(I,\mathcal{E})$ by drawing the corresponding wires dotted):
	\[
		\raisebox{1.25cm}{
			\scalebox{\picturescaling}{$
				\raisebox{2cm}{$(u,\sigma)$}
				\hspace{5mm}\raisebox{2cm}{$=$}
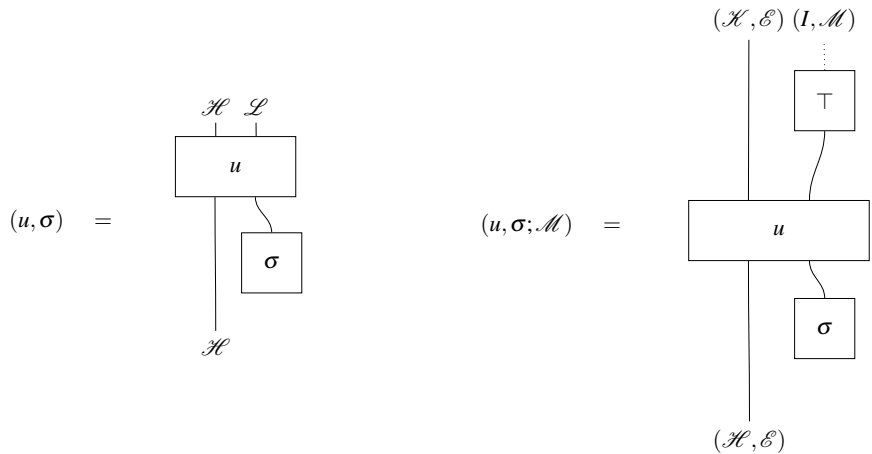
\hspace{5mm}
				\begin{tikzpicture}
	\path [use as bounding box] (-1.50cm,-2.00cm) -- (1.50cm,-2.00cm) -- (1.50cm,2.00cm) -- (-1.50cm,2.00cm) -- (-1.50cm,-2.00cm);
	\begin{pgfonlayer}{nodelayer}
		\node (0) at (-0.28,-2.00) {$\mathcal{H}$};
		\node (1) at (-0.28,2.00) {$\mathcal{H}$};
		\node (2) at (0.39,2.00) {$\mathcal{L}$};
		\node [style=box] (3) at (0.65,-0.60) {$\sigma$};
		\node [style=widebox] (4) at (0.05,1.00) {$u$};
	\end{pgfonlayer}
	\begin{pgfonlayer}{edgelayer}
		\draw [out=90.00,in=270.00] (4.123.69) to (1);
		\draw [out=90.00,in=270.00] (4.56.31) to (2);
		\draw [out=90.00,in=270.00] (0) to (4.236.31);
		\draw [out=90.00,in=270.00] (3.90.00) to (4.303.69);
	\end{pgfonlayer}
\end{tikzpicture}
			$}
		}
		\hspace{2cm}
		\scalebox{\picturescaling}{$
			\raisebox{3.5cm}{$(u,\sigma;\mathcal{M})$}
			\hspace{5mm}\raisebox{3.5cm}{$=$}\hspace{5mm}
			\begin{tikzpicture}
	\path [use as bounding box] (-2.00cm,-3.50cm) -- (2.00cm,-3.50cm) -- (2.00cm,3.50cm) -- (-2.00cm,3.50cm) -- (-2.00cm,-3.50cm);
	\begin{pgfonlayer}{nodelayer}
		\node (0) at (-0.37,-3.50) {$(\mathcal{H},\mathcal{E})$};
		\node (1) at (-0.38,3.50) {$(\mathcal{K},\mathcal{E})$};
		\node (2) at (0.87,3.50) {$(I,\mathcal{M})$};
		\node [style=box] (3) at (0.88,-1.63) {$\sigma$};
		\node [style=widebox,minimum width=3cm] (4) at (0.12,-0.00) {$u$};
		\node [style=box] (5) at (0.88,2.16) {$\top$};
	\end{pgfonlayer}
	\begin{pgfonlayer}{edgelayer}
		\draw [dotted,out=90.00,in=270.00] (5.90.00) to (2);
		\draw [out=90.00,in=270.00] (0) to (4.225.00);
		\draw [out=90.00,in=270.00] (3.90.00) to (4.315.00);
		\draw [out=90.00,in=270.00] (4.135.00) to (1);
		\draw [out=90.00,in=270.00] (4.45.00) to (5.270.00);
	\end{pgfonlayer}
\end{tikzpicture}
		$}
	\]
	The diagrammatic representation is justified, in hindsight, by the fact that these are morphisms in a symmetric partially-monoidal category. It could not be included in the main body of the theorems because of space constraints. Composition of processes is defined as follows (pure processes on the left, general processes on the right):
	\[
		\raisebox{1.75cm}{
		\scalebox{\picturescaling}{$
			\raisebox{3.25cm}{$(v,\tau) \circ (u,\sigma)$}
			\hspace{5mm}\raisebox{3.25cm}{$=$}
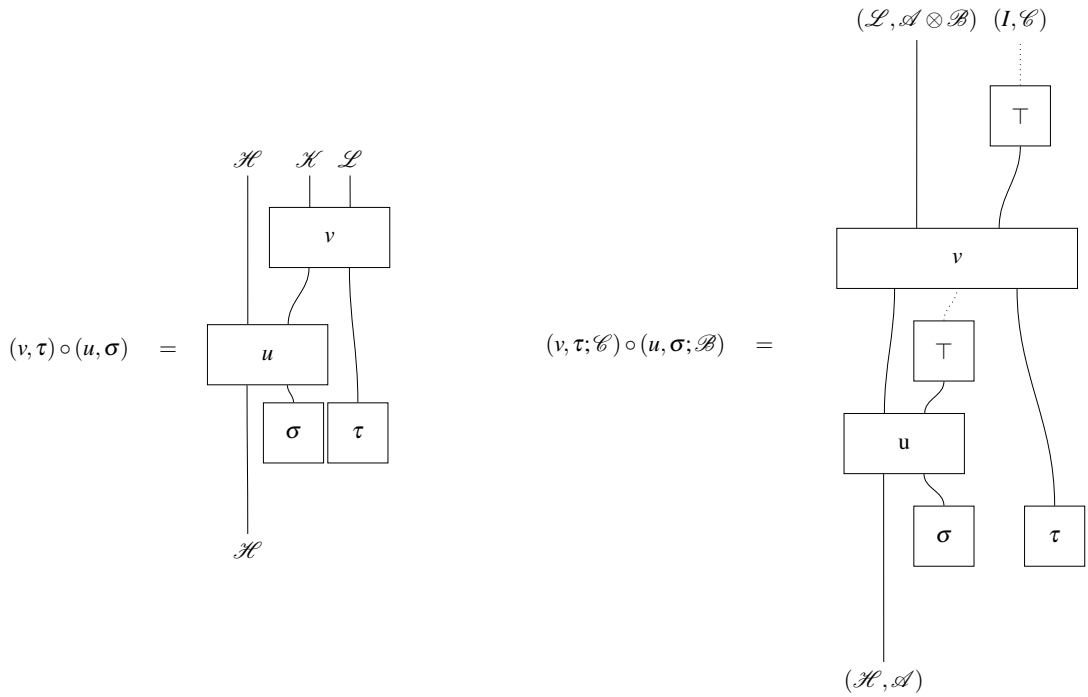
\hspace{5mm}
			\begin{tikzpicture}
	\path [use as bounding box] (-1.50cm,-3.25cm) -- (1.50cm,-3.25cm) -- (1.50cm,3.25cm) -- (-1.50cm,3.25cm) -- (-1.50cm,-3.25cm);
	\begin{pgfonlayer}{nodelayer}
		\node (0) at (-0.83,-3.25) {$\mathcal{H}$};
		\node (1) at (-0.83,3.25) {$\mathcal{H}$};
		\node (2) at (0.20,3.25) {$\mathcal{K}$};
		\node (3) at (0.87,3.25) {$\mathcal{L}$};
		\node [style=box] (4) at (-0.07,-1.30) {$\sigma$};
		\node [style=box] (5) at (1.00,-1.30) {$\tau$};
		\node [style=widebox] (6) at (-0.50,-0.00) {$u$};
		\node [style=widebox] (7) at (0.53,1.95) {$v$};
	\end{pgfonlayer}
	\begin{pgfonlayer}{edgelayer}
		\draw [out=90.00,in=270.00] (7.123.69) to (2);
		\draw [out=90.00,in=270.00] (7.56.31) to (3);
		\draw [out=90.00,in=270.00] (0) to (6.236.31);
		\draw [out=90.00,in=270.00] (4.90.00) to (6.303.69);
		\draw [out=90.00,in=270.00] (6.123.69) to (1);
		\draw [out=90.00,in=270.00] (6.56.31) to (7.236.31);
		\draw [out=90.00,in=270.00] (5.90.00) to (7.303.69);
	\end{pgfonlayer}
\end{tikzpicture}
		$}
		}
		\hspace{2cm}
		\scalebox{\picturescaling}{$
			\raisebox{5.5cm}{$(v,\tau;\mathcal{C}) \circ (u,\sigma;\mathcal{B}) $}
			\hspace{5mm}\raisebox{5.5cm}{$=$}\hspace{5mm}
			\begin{tikzpicture}
	\path [use as bounding box] (-2.50cm,-5.50cm) -- (2.50cm,-5.50cm) -- (2.50cm,5.50cm) -- (-2.50cm,5.50cm) -- (-2.50cm,-5.50cm);
	\begin{pgfonlayer}{nodelayer}
		\node (0) at (-1.17,-5.50) {$(\mathcal{H},\mathcal{A})$};
		\node (1) at (-0.62,5.50) {$(\mathcal{L},\mathcal{A}\otimes\mathcal{B})$};
		\node (2) at (1.10,5.50) {$(I,\mathcal{C})$};
		\node [style=box] (3) at (-0.17,-3.08) {$\sigma$};
		\node [style=box] (4) at (1.67,-3.08) {$\tau$};
		\node [style=widebox] (5) at (-0.83,-1.54) {u};
		\node [style=box] (6) at (-0.17,0.00) {$\top$};
		\node [style=widebox,minimum width=4cm] (7) at (0.05,1.54) {$v$};
		\node [style=box] (8) at (1.10,3.91) {$\top$};
	\end{pgfonlayer}
	\begin{pgfonlayer}{edgelayer}
		\draw [dotted,out=90.00,in=270.00] (8.90.00) to (2);
		\draw [out=90.00,in=270.00] (0) to (5.236.31);
		\draw [out=90.00,in=270.00] (3.90.00) to (5.303.69);
		\draw [out=90.00,in=270.00] (5.56.31) to (6.270.00);
		\draw [dotted,out=90.00,in=270.00] (6.90.00) to (7.270.00);
		\draw [out=90.00,in=270.00] (7.143.13) to (1);
		\draw [out=90.00,in=270.00] (7.36.87) to (8.270.00);
		\draw [out=90.00,in=270.00] (5.123.69) to (7.206.57);
		\draw [out=90.00,in=270.00] (4.90.00) to (7.333.43);
	\end{pgfonlayer}
\end{tikzpicture}
		$}
	\]
	Tensor product of processes is defined as follows (pure processes above, general processes below):
	\[
		\scalebox{\picturescaling}{$
			\raisebox{2.25cm}{$(u,\sigma) \otimes (v,\tau) $}
			\hspace{5mm}\raisebox{2.25cm}{$=$}
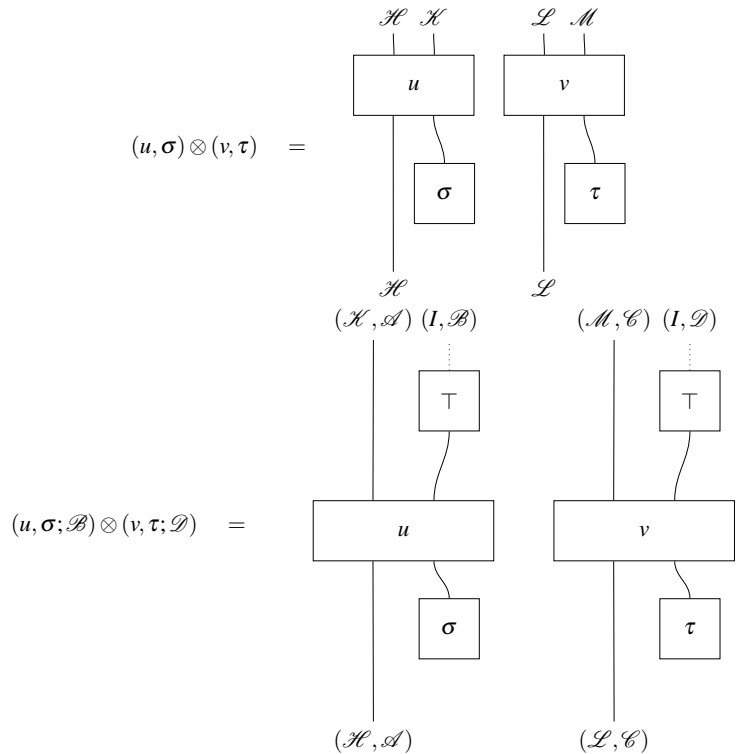
\hspace{5mm}
			\begin{tikzpicture}
	\path [use as bounding box] (-2.50cm,-2.25cm) -- (2.50cm,-2.25cm) -- (2.50cm,2.25cm) -- (-2.50cm,2.25cm) -- (-2.50cm,-2.25cm);
	\begin{pgfonlayer}{nodelayer}
		\node (0) at (-1.55,-2.25) {$\mathcal{H}$};
		\node (1) at (0.95,-2.25) {$\mathcal{L}$};
		\node (2) at (-1.55,2.25) {$\mathcal{H}$};
		\node (3) at (-0.88,2.25) {$\mathcal{K}$};
		\node (4) at (0.95,2.25) {$\mathcal{L}$};
		\node (5) at (1.62,2.25) {$\mathcal{M}$};
		\node [style=box] (6) at (-0.70,-0.68) {$\sigma$};
		\node [style=box] (7) at (1.80,-0.68) {$\tau$};
		\node [style=widebox] (8) at (-1.21,1.12) {$u$};
		\node [style=widebox] (9) at (1.29,1.12) {$v$};
	\end{pgfonlayer}
	\begin{pgfonlayer}{edgelayer}
		\draw [out=90.00,in=270.00] (8.123.69) to (2);
		\draw [out=90.00,in=270.00] (8.56.31) to (3);
		\draw [out=90.00,in=270.00] (9.123.69) to (4);
		\draw [out=90.00,in=270.00] (9.56.31) to (5);
		\draw [out=90.00,in=270.00] (0) to (8.236.31);
		\draw [out=90.00,in=270.00] (6.90.00) to (8.303.69);
		\draw [out=90.00,in=270.00] (1) to (9.236.31);
		\draw [out=90.00,in=270.00] (7.90.00) to (9.303.69);
	\end{pgfonlayer}
\end{tikzpicture}
		$}
	\]
	\[
		\scalebox{\picturescaling}{$
			\raisebox{3.5cm}{$(u,\sigma;\mathcal{B}) \otimes (v,\tau;\mathcal{D}) $}
			\hspace{5mm}\raisebox{3.5cm}{$=$}\hspace{5mm}
			\begin{tikzpicture}
	\path [use as bounding box] (-4.00cm,-3.50cm) -- (4.00cm,-3.50cm) -- (4.00cm,3.50cm) -- (-4.00cm,3.50cm) -- (-4.00cm,-3.50cm);
	\begin{pgfonlayer}{nodelayer}
		\node (0) at (-2.38,-3.50) {$(\mathcal{H},\mathcal{A})$};
		\node (1) at (1.62,-3.50) {$(\mathcal{L},\mathcal{C})$};
		\node (2) at (-2.38,3.50) {$(\mathcal{K},\mathcal{A})$};
		\node (3) at (-1.12,3.50) {$(I,\mathcal{B})$};
		\node (4) at (1.62,3.50) {$(\mathcal{M},\mathcal{C})$};
		\node (5) at (2.88,3.50) {$(I,\mathcal{D})$};
		\node [style=box] (6) at (-1.12,-1.63) {$\sigma$};
		\node [style=widebox,minimum width=3cm] (7) at (-1.88,-0.00) {$u$};
		\node [style=box] (8) at (-1.12,2.16) {$\top$};
		\node [style=box] (9) at (2.88,-1.63) {$\tau$};
		\node [style=widebox,minimum width=3cm] (10) at (2.12,-0.00) {$v$};
		\node [style=box] (11) at (2.88,2.16) {$\top$};
	\end{pgfonlayer}
	\begin{pgfonlayer}{edgelayer}
		\draw [dotted,out=90.00,in=270.00] (8.90.00) to (3);
		\draw [out=90.00,in=270.00] (0) to (7.225.00);
		\draw [out=90.00,in=270.00] (6.90.00) to (7.315.00);
		\draw [out=90.00,in=270.00] (7.135.00) to (2);
		\draw [out=90.00,in=270.00] (7.45.00) to (8.270.00);
		\draw [dotted,out=90.00,in=270.00] (11.90.00) to (5);
		\draw [out=90.00,in=270.00] (1) to (10.225.00);
		\draw [out=90.00,in=270.00] (9.90.00) to (10.315.00);
		\draw [out=90.00,in=270.00] (10.135.00) to (4);
		\draw [out=90.00,in=270.00] (10.45.00) to (11.270.00);
	\end{pgfonlayer}
\end{tikzpicture}
		$}
	\]
	The final thing to note, specifically for the proof of theorem \ref{theorem:process-theory}, is that the operation of restriction $\Restr{\mathcal{K}}{u(\rho \otimes \sigma)}$ in the definition of processes is really the same as applying the tensor product of the identity on $(\mathcal{K},\mathcal{E})$ with the discarding map $\top: (\mathcal{M},I) \rightarrow (I,\mathcal{M})$. This is true because whenever $\rho$ is a (not necessarily pure) state for $\mathcal{K} \otimes \mathcal{M}$, the restriction of $\rho$ to $\mathcal{K}$ can be done equivalently by taking $\Transf{\mathcal{K}}' \rho$ or by taking $\Transf{\mathcal{M}} \rho$, because of the orthocomplementarity requirement involved in the definition of the tensor product $\mathcal{K} \otimes \mathcal{M}$.
\end{proof}

\subsubsection*{Proposition \ref{proposition:purification-principle}}
\begin{proof}
	The isomorphism is straightforward, as is the observation that transformations + pure states generate $\ProcessTheoryPure{\Xi}{\Theta}$  (by definition of morphisms in $\ProcessTheoryPure{\Xi}{\Theta}$ ) and that the addition of discarding maps generates $\ProcessTheory{\Xi}{\Theta}$ (by definition of morphisms in $\ProcessTheory{\Xi}{\Theta}$). Then only thing to be careful about is the meaning of ``smallest'' symmetric partially-monoidal category. Because of the generators chosen, any such subcategory necessarily contains all objects of $\ProcessTheory{\Xi}{\Theta}$: because the tensor product on the subcategory is the one inherited from $\ProcessTheory{\Xi}{\Theta}$, then the subcategory also has the same tensor products defined that $\ProcessTheory{\Xi}{\Theta}$ has.
\end{proof}

\subsubsection*{Proposition \ref{proposition:causality}}
\begin{proof}
	A discussed in the final paragraph of the proof to theorems \ref{theorem:pure-process-theory} and \ref{theorem:process-theory}, the operation of restriction $\Restr{\mathcal{K}}{u(\rho \otimes \sigma)}$ in the definition of processes is really the same as applying the tensor product of the identity on $(\mathcal{K},\mathcal{E})$ with the discarding map $\top: (\mathcal{M},I) \rightarrow (I,\mathcal{M})$. This, together with proposition \ref{proposition:purification-principle}, makes this statement self-evident.
\end{proof}

\end{document}